\documentclass[conference]{IEEEtran}
\IEEEoverridecommandlockouts
\usepackage{cite}
\usepackage{amsmath,amssymb,amsfonts}
\usepackage{algorithmic}
\usepackage{graphicx}
\usepackage{multirow}
\usepackage{textcomp}
\usepackage{xcolor}
\usepackage{url}
\usepackage[inkscapelatex=false]{svg}
\usepackage[a4paper, total={184mm,239mm}]{geometry}
\def\BibTeX{{\rm B\kern-.05em{\sc i\kern-.025em b}\kern-.08em
    T\kern-.1667em\lower.7ex\hbox{E}\kern-.125emX}}
\begin{document}

\title{Fast and Low-Cost Approximate Multiplier for FPGAs using Dynamic Reconfiguration}

\author{\IEEEauthorblockN{Shervin Vakili\IEEEauthorrefmark{1},
Mobin Vaziri\IEEEauthorrefmark{2}, Amirhossein Zarei\IEEEauthorrefmark{1} and
J.M. Pierre Langlois\IEEEauthorrefmark{2}}
\IEEEauthorblockA{\IEEEauthorrefmark{1} Institut national de la recherche scientifique (INRS), Montréal, Canada\\
\IEEEauthorrefmark{2}Department of Computer and Software Engineering, Polytechnique Montréal, Montréal, Canada\\
\{shervin.vakili, amirhossein.zarei\}@inrs.ca
\{mobin.vaziri, pierre.langlois\}@polymtl.ca
}}

\maketitle

\begin{abstract}
Multipliers are widely-used arithmetic operators in digital signal processing and machine learning circuits. Due to their relatively high complexity, they can have high latency and be a significant source of power consumption. One strategy to alleviate these limitations is to use approximate computing. This paper thus introduces an original FPGA-based approximate multiplier specifically optimized for machine learning computations. It utilizes dynamically reconfigurable lookup table (LUT) primitives in AMD-Xilinx technology to realize the core part of the computations. The paper provides an in-depth analysis of the hardware architecture, implementation outcomes, and accuracy evaluations of the multiplier proposed in INT8 precision. Implementation results on an AMD-Xilinx Kintex Ultrascale+ FPGA demonstrate remarkable savings of 64\% and 67\% in LUT utilization for signed multiplication and multiply-and-accumulation configurations, respectively, when compared to the standard Xilinx multiplier core.
Accuracy measurements on four popular deep learning (DL) benchmarks indicate a minimal average accuracy decrease of less than 0.29\% during post-training deployment, with the maximum reduction staying less than 0.33\%. The source code of this work is available on GitHub\footnote{\url{https://github.com/INRS-ECCoLe/DyRecMul}}.
\end{abstract}

\vspace{2mm}


\section{Introduction \& Related Works}\label{1}
Computational circuits may become a bottleneck in wireless communications, digital signal processing, multimedia, image processing, machine learning (ML), etc., due to their high energy consumption, long delay, and large circuit area utilization. Applications such as neural networks have inherent error resilience, which presents an opportunity to use approximation techniques for enhancing their efficiency \cite{jiang2020approximate, chippa2013analysis}. Multiplication is a fundamental operation in computations, and, to enhance its efficiency, various approximate multipliers have been proposed. These multipliers aim to reduce delay, energy consumption and area \cite{10.1145/3610291}. In 1962, Mitchell \cite{mitchell1962computer} introduced the first logarithm-based multiplier. This multiplier converted multiplications into addition operations, and it deployed approximation methods to calculate logarithmic expressions. Since then, there have been significant advancements in logarithm-based approximate multipliers \cite{liu2018design,ansari2020improved,ansari2019hardware,saadat2018minimally}. Liu et al. used approximate adders to enhance accuracy and reduce power consumption in logarithm-based multipliers \cite{liu2018design}. Given that multiplication is a nonlinear operation, several works proposed methods to use linearization for approximate multiplication \cite{imani2019approxlp,chen2020optimally}. An approximate linearization algorithm, called ApproxLP \cite{imani2019approxlp}, utilized comparators to separate different sub-domains and allocate a proper linear function to each one instead of a nonlinear multiplication. This method, however, requires additional comparators to increase accuracy and reduce sub-domain size, which can result in longer delays and increased area \cite{10.1145/3610291}. Chen et al. introduced an optimally approximate multiplier (OAM) \cite{chen2020optimally} to improve linearization and minimize the number of comparators, resulting in lower delay and improved performance. Other studies suggest utilizing a combination of accurate and approximate multipliers with varying precision levels to achieve the necessary accuracy, but this may come at the cost of higher energy consumption and circuit area \cite{imani2017cfpu,imani2018rmac}.

In non-logarithmic multiplication, the approximation can be introduced at four distinct stages: data input, partial product generation, accumulation, and Booth encoding \cite{10.1145/3610291}. Truncating input data is often used to reduce bitwidth, and it can be done through methods such as the dynamic segment method (DSM) and static segment method (SSM) \cite{narayanamoorthy2014energy,vahdat2019tosam,hashemi2015drum,vahdat2017letam}. DSM keeps a specified number of bits from the most-significant '1' in an $n$-bit operand but requires more hardware resources than SSM. SSM has limited and pre-determined options for truncating input data, resulting in more redundant bits \cite{10.1145/3610291, narayanamoorthy2014energy}. Some papers have introduced approximations in the partial product generation stage \cite{kulkarni2011trading, venkatachalam2017design,yang2018low, yang2017low}. However, the most beneficial stage to introduce approximation is during partial product accumulation, which typically requires the largest circuit area and computation time.

Compressors can be used for counting ones in tree-based partial product accumulations \cite{10.1145/3610291}, such as the Wallace tree \cite{wallace1964suggestion} and Dadda tree \cite{dadda1965some}. Instead of exact compressors, a variety of approximate compressors have been introduced to decrease circuit area and delay \cite{venkatachalam2017design,yang2018low,esposito2018approximate,8017557,tung2019low,yang2015approximate,momeni2014design,mahdiani2009bio,lin2013high,akbari2017dual,ahmadinejad2019energy,strollo2020comparison,marimuthu2016design,wang2022minac}. These include the approximate 1-bit half adder, approximate 1-bit full adder, and approximate 4-2 compressors \cite{yang2015approximate,momeni2014design,akbari2017dual,strollo2020comparison}. To further decrease energy consumption, delay, and hardware utilization, high-order approximate compressors with more than five inputs have been proposed \cite{esposito2018approximate,tung2019low,ahmadinejad2019energy,marimuthu2016design}. Mahdiani et al. proposed a technique that removes the least significant partial products from the partial product matrix to decrease circuit area and delay, although it comes with some degree of error which can be adjusted by modifying the truncation level \cite{mahdiani2009bio}.

The existing approximate methods, which have been mainly designed to reduce energy consumption and area utilization in ASIC implementations, may not be as effective in FPGAs. This is because FPGA reconfigurable logic fabrics are typically based on fixed-size look-up tables (LUTs). Modern AMD-Xilinx and Intel FPGAs have hardwired DSP multipliers that offer faster and more energy-efficient multiplication than soft implementation on general-purpose LUT fabrics. The DSP-based multipliers are valuable resources that are situated in specific locations, which can result in long routing delays. On the other hand, LUTs are spread out across the chip, making them more easily routable. Moreover, there is a limitation on the number of DSP-based multipliers available. One solution is to deploy approximation methods to enhance the efficiency of LUT-based multiplication, in terms of speed, energy/power consumption and hardware utilization. Ullah et al. \cite{ullah2018area,ullah2018smapproxlib} implemented an optimization technique by truncating a least significant partial product of a 4×2 multiplier to reduce LUT utilization. Van Toan et al. \cite{van2020fpga} designed compact 3-2 and 4-2 compressors for use in different approximate multipliers with varying levels of accuracy. Kumm et al. \cite{kumm2013dynamically} proposed dynamically reconfigurable FIR filters in Xilinx FPGAs using configurable look-up tables (CFGLUTs).

This paper introduces DyRecMul, a cost-effective dynamically reconfigurable approximate multiplier for FPGAs. It is optimized for machine learning computation, has a short critical path, and uses a small number of LUTs. DyRecMul utilizes AMD-Xilinx technology's reconfigurable LUT primitives to approximate multiplication without significant accuracy degradation, even in post-training inference. To preserve dynamic range, DyRecMul utilizes a cost-effective encoder that transforms a fixed-point operand into an 8-bit floating-point format, and a decoder to revert the result back to fixed-point. The paper presents the design details and evaluation results of an INT8 version of DyRecMul since INT8 is a popular datatype in cost-effective machine learning computing.  The multiplier introduces negligible accuracy loss while reducing significantly the number of required LUTs in FPGAs compared to the standard exact multiplier. Additionally, DyRecMul boasts very low latency, which allows for a faster clock frequency than that of typical AMD-Xilinx multiplier cores. The key contributions of this paper are summarized below.

\begin{itemize}

    \item Utilization of dynamically reconfigurable LUTs (CFGLUTs) in AMD-Xilinx FPGAs to make a low-cost and fast short-bitwidth multiplier. 
    \item Internal conversion of fixed-point to a floating-point format to preserve dynamic range that is beneficial for machine learning and deep learning applications.
    \item For an INT8 case study, illustration of the design detail of a highly optimized encoder circuit to convert INT8 to 8-bit floating point format and a low-cost decoder to convert 8-bit floating point format to INT8.
    
\end{itemize}

The paper is organized as follows. Section \ref{2} presents the analytical foundation of the proposed approach and illustrates the microarchitectural design detail of DyRecMul for the INT8 case study. Section \ref{3} provides the evaluation results including error analysis, hardware evaluations and accuracy measurements for deep learning applications. Finally, Section \ref{4} concludes the paper.

\section{Proposed Approximate Multiplier}\label{2}
This section provides an overview of the analytical basis of DyRecMul, followed by a detailed description of an INT8 DyRecMul design which also provides insight into the design decisions and considerations behind them.

\subsection{Analytical description}\label{2.1}
A precise \textit{N}-bit signed integer multiplication of two operands, \textit{X} and \textit{W}, can be represented by:

\begin{equation}\label{eq:1}
    Z=s_z.\sum_{i=0}^{2N-2}{z_i.2^i}=\ s_x.\sum_{i=0}^{N-2}{x_i.2^i}\ \times\ s_w.\sum_{i=0}^{N-2}{w_i.2^i}  
\end{equation}
where $x_i$, $w_i$ and $z_i$ denote the $i^{th}$ bit of operand $X$ and $W$ and the result $Z$, respectively. $S_x$ takes the value of -1 when X is negative, and 0 otherwise. The same logic applies to $S_z$ and $S_y$. In INT8 representation, $N=8$. When using multipliers in a computing system that only supports a single datatype, the output $Z$ must be expressed in the same format as the input operands. This can be done using techniques like truncation.

The proposed approach utilizes an encoder to convert the first operand, $X$, to a floating-point representation, ${\hat{X}}_{float}$, of format

\begin{equation}\label{eq:2}
    \begin{aligned}
        &float\left({sign}_{BW},{exp}_{BW},{mnt}_{BW}\right),\\ 
        &{exp}_{BW}>0,\ {mnt}_{BW}<N-1  
    \end{aligned}
\end{equation}
where ${sign}_{BW}$, ${exp}_{BW}$ and ${mnt}_{BW}$ denote the bitwidth of the sign, exponent and mantissa elements, respectively. Since we assume a signed multiplication, ${sign}_{BW} =1$, allocating a single bit to the sign. To cover the entire dynamic range of $X$, ${exp}_{BW}$ and ${mnt}_{BW}$ must fulfill the following:

\begin{equation}\label{eq:3}
    {exp}_{BW}+{mnt}_{BW}>N-1
\end{equation}
The mantissa is a segment of ${mnt}_{BW}$ bits from $\left|X\right|$, with the leftmost bit in $\left|X\right|$ that contains '1' being the most significant bit. Since the mantissa is shorter than the magnitude bits of $X$, this conversion involves an approximation. The conversion can be expressed as:

\begin{equation}\label{eq:4}
{\hat{X}}_{float}=\ \left[{\hat{X}}_{sign},\ {\hat{X}}_{exp}\ ,\ {\hat{X}}_{mnt}\ \right]
\end{equation}

\[
\begin{cases}
    {\hat{X}}_{sign} = x_{N - 1},&\\
    {\hat{X}}_{exp} = N - i ~\textit{where} \begin{cases}
                                                2^{i} < \left|X\right| \leq 2^{i - 1},~X \geq 2^{mnt_{BW}} &\\
                                                0,~X < 2^{mnt_{BW}},
                                            \end{cases}& \\
    {\hat{X}}_{mnt} = \textit{Round}~(\frac{\left|X\right|}{2^{exp}})
\end{cases}
\]

The mantissa ${\hat{X}}_{mnt}$ is then multiplied by $\left|W\right|$ to generate the mantissa of the result. To keep the result in the format of Eq. (\ref{eq:2}), the product is quantized to ${mnt}_{BW}$ bits using:

\begin{equation}\label{eq:5}
    {\hat{Z}}_{mnt}=Q\left({\hat{X}}_{mnt}\times\left|W\right|,\ {mnt}_{BW}\right),
\end{equation}
where $Q$ is the quantization function. The result must be converted back to integer format. For this purpose, the $(N-1)$-bit absolute value of $Z$ is first calculated by:

\begin{equation}\label{eq:6}
    {\left|Z\right|=2^{{\hat{X}}_{exp}}\times\hat{Z}}_{mnt}
\end{equation}
The result, $Z$, in integer format is obtained by applying a two's complement function when $Z$ is negative. The sign of $Z$ is determined by XORing the sign bits of $X$ and $W$.

\begin{equation}\label{eq:7}
    Z = S_{z}.\left|Z\right|,  
    \begin{cases}
        S_{z} = +1,~x_{N-1} \bigoplus w_{N-1} = 0 &\\
        S_{z} = -1,~x_{N-1} \bigoplus w_{N-1} = 1
    \end{cases}
\end{equation}

\subsection{DyRecMul for INT8 multiplication: architecture and components}\label{2.2}
\begin{figure}
    \centering
    \includegraphics[width=\columnwidth]{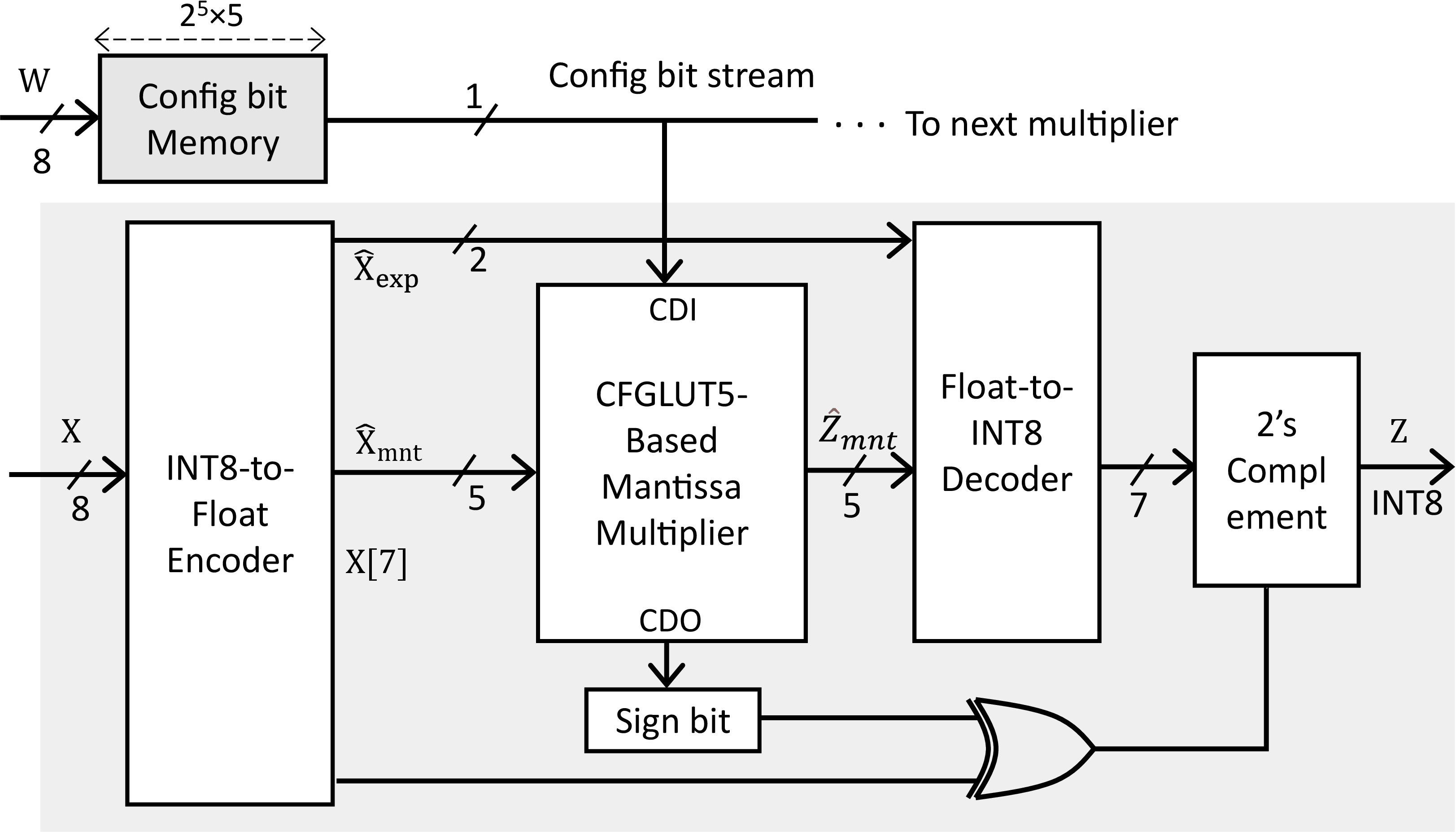}
    \caption{Architecture of DyRecMul.}
    \label{fig:1}
\end{figure}

INT8 quantization is a supported feature in machine learning frameworks such as TensorFlow Lite and PyTorch, as well as hardware toolchains like AMD-Xilinx DNNDK. Moreover, INT8 is a popular datatype in machine learning hardware accelerators and ML-optimized GPUs designed for embedded and edge applications \cite{lin2020int8}, \cite{kim2021int8}. The INT8 DyRecMul described in this section can be used in place of INT8 multipliers for pre- or post-training inference. This multiplier is intended primarily for single-datatype INT8 architectures, meaning that it calculates $Z=X \times W$, where $X$, $W$ and $Z$ are all INT8. Fig. \ref{fig:1} depicts its architecture, which consists of four main components: (1) a cost-effective INT8 to floating-point encoder; (2) an ultra-low-cost mantissa multiplier using dynamically reconfigurable LUTs; (3) a floating-point to INT8 decoder; (4) a two’s complement logic. The following subsections describe each component in detail.

\textit{A) Integer to floating-point encoder—} The proposed architecture converts its first operand, $X$, from INT8 to $float\left(1,2,5\right)$ representation, where $1$, $2$, and $5$ indicate the number of allocated bits to the sign, exponent and mantissa, respectively. This conversion corresponds to Eq. (\ref{eq:4}) in Section \ref{2.1}. The mantissas are multiplied and the exponents are added, and this conversion limits the binary multiplication to five bits while maintaining the dynamic range, which is crucial for accurate DL computations. As will be discussed later, DyRecMul deploys a CGFLUT5-based unsigned multiplier in which one operand must be five bits wide to achieve optimal efficiency. When this multiplier is used in a weight stationary DL accelerator, input feature maps are fed as the first operand, $X$, to the multipliers. Without a floating-point conversion, the feature maps would need to be directly quantized to five bits. That would significantly limit the supported dynamic range, which could cause important accuracy loss for small activation values and ultimately lead to a prohibitive level of inaccuracy in machine learning inference. More precisely, such a quantization would limit the range of supported values to $[-2^{5}, 2^{5}-1]$, while $float\left(1,2,5\right)$ expands the range to $[-2^{3}\times2^{5}, 2^{3}\times2^{5}-1]$ and demands the same unsigned multiplier size for mantissa multiplication. The experimental results in Section \ref{3.3} demonstrate that this conversion greatly helps in maintaining the precision of DL calculations.

\begin{figure*}
    \centering
    \includegraphics[width = 420pt]{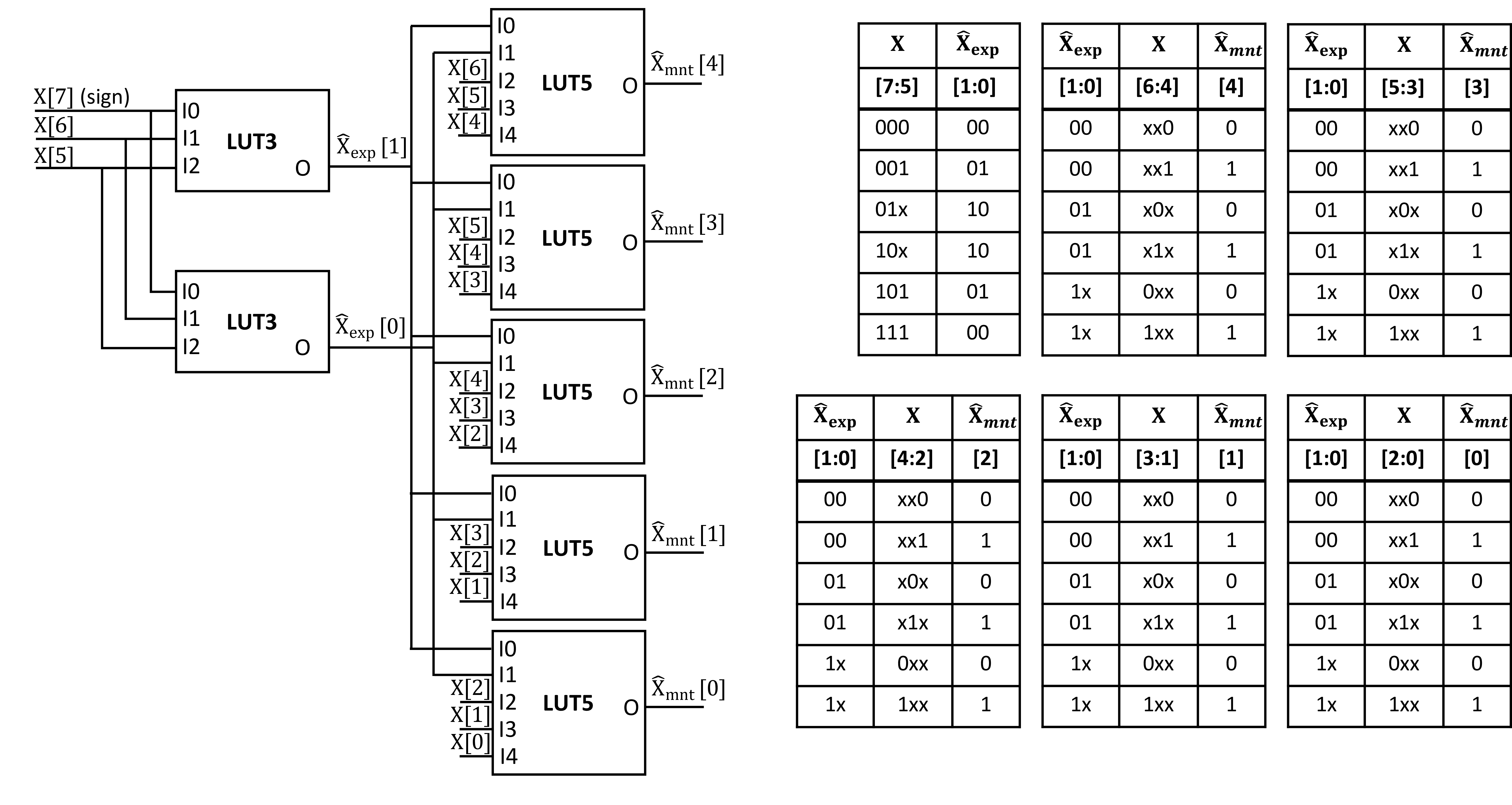}
    \caption{INT8 to $float\left(1,2,5\right)$ encoder, LUT mapping and corresponding truth tables.}
    \label{fig:2}
\end{figure*}

The conversion logic from INT8 to $float\left(1,2,5\right)$ is designed to be low-cost and efficient, requiring only seven LUT5 elements. Fig. \ref{fig:2} depicts the truth tables and the corresponding LUT5 allocations for this encoder. The $2$-bit exponent is acquired from the sign bit and the two most significant bits, while each mantissa bit is obtained from the exponent and three corresponding input bits. The encoder's critical path consists of two LUT5 units and their corresponding routing circuits.

\textit{B) Dynamically reconfigurable mantissa multiplier—} The core component of DyRecMul is a low-cost unsigned integer multiplier that relies on AMD-Xilinx CFGLUT5 primitives. As shown in Fig. \ref{fig:1}, this component implements mantissa multiplication according to Eq. (\ref{eq:5}). Instead of implementing the logic circuit of an ordinary double-operand multiplier, this approach translates the second operand into the LUT primitive configuration bits. The first operand is given as input to the LUTs. When the second operand changes, the LUTs have to be reconfigured. Due to the latency overheads associated with reconfiguration, DyRecMul is best suited for applications where one of the operands does not change frequently. Several deep learning hardware acceleration dataflows and digital signal processing systems possess this attribute.

\begin{figure*}[h]
    \centering
    \includegraphics[width = 270pt]{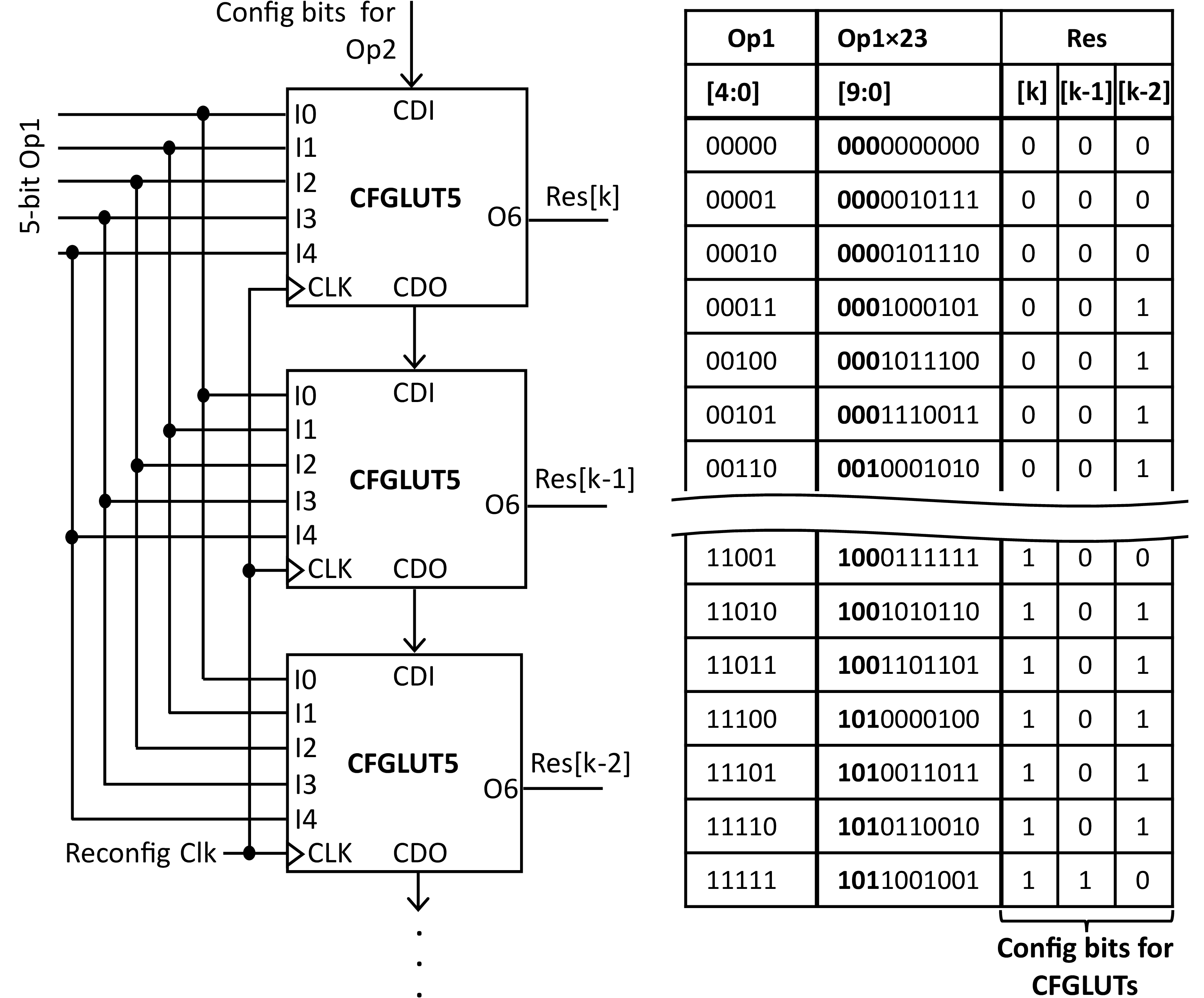}
    \caption{$5$-bit dynamically reconfigurable multiplier with $3$-bit result using CFGLUT5 primitives. The second operand, $Op2=23$ in this example.}
    \label{fig:3}
\end{figure*}

The configurable LUT primitives in AMD-Xilinx FPGAs, CFGLUT5, have five input bits, $2^{5}$ bits of lookup entries and one-bit output. Fig. \ref{fig:3} illustrates the LUT mapping and configuration bits for an example $5$-bit unsigned multiplier. The output consists of the quantized $k$ most significant bits of the result. This means that only $k$ CFGLUT5s are needed, one for each output bit. As an example, this figure assumes a value of $23$ for the second operand, $Op2$, meaning that CFGLUT5s are configured to generate the product of $5$-bit quantized first operand, $Op1$, and $Op2=23$. Fig. 3 depicts the process of calculating the configuration bits for three CFGLUT5s which produce the three most significant bits of the product. Rounding is used in these calculations to minimize quantization error. The configuration bits are loaded into the CFGLUTs serially through a cascaded CDI and CDO chain. As the bit width of the first operand surpasses five, the number of CFGLUT5s required for each output bit grows exponentially. More precisely, the required number of CFGLUT5s is:

\begin{equation}\label{eq:8}
    \#CFGLUT5=2^{b1-5}\times k
\end{equation}
where $b1$ and $k$ denote the bitwidth of the first operand, $Op1$, and the result, respectively. 

As an example, implementing an exact $8\times8$ unsigned multiplier requires $2^{3}\times16=128$ CFGLUT5s. Although the LUT utilization may appear to be high, a more serious obstacle lies in the significant number of reconfiguration bits - a total of $4096 (128\times25)$. In order to fully exploit the CFGLUT5s while minimizing their quantity, we limited $Op1$ to five bits resulting in one CFGLUT utilization per result bit. Thus, using $k$ parallel CFGLUT5s allows for calculating the $k$-bit result of a $5$-bit multiplication. If the bitwidth is shorter, some CFGLUTs may remain partially unused, while increasing it to over five bits will exponentially increase the number of CFGLUT5s. Additionally, for the INT8 DyRecMul, we set the result bitwidth $k$ to five, restricting the number of CFGLUT5s to only five units and the number of reconfiguration bits to $160$. When this multiplier is used in a weight stationary DL accelerator, weights are translated into the configuration bits in CFGLUT5s and input feature maps are fed as operands to the multipliers. As there is no interdependence between CFGLUT5 elements, they can operate entirely in parallel, resulting in a short critical path of just one CFGLUT5.

The configuration bits for CFGLUT5s can be stored in shared SRAM-based memory, as shown in Fig. \ref{fig:1}. Whenever the second operand $W$ changes, the corresponding $160$-bit configuration sequence for the new $W$ is retrieved from this memory and written to CFGLUTs via the cascaded CDI/CDO chain in a serial manner.  The configuration bit memory can be shared among multiple multipliers and fits within one BRAM36 unit. The reconfiguration time is an important factor in the efficiency of DyRecMul. As previously mentioned, DyRecMul is most suitable for applications where one of the operands does not change frequently. In other words, DyRecMac gains an edge when one of the operands remains unchanged significantly longer than the reconfiguration time. The influence of reconfiguration time on the overall processing time primarily relies on the dataflow and architecture of the DL accelerator. For example, within a "more optimal" weight-stationary architecture, weights are extensively reused once loaded into the processing elements, thereby minimizing the frequency of reconfiguration.

\begin{figure*}[h]
    \centering
    \includegraphics[width = 500pt]{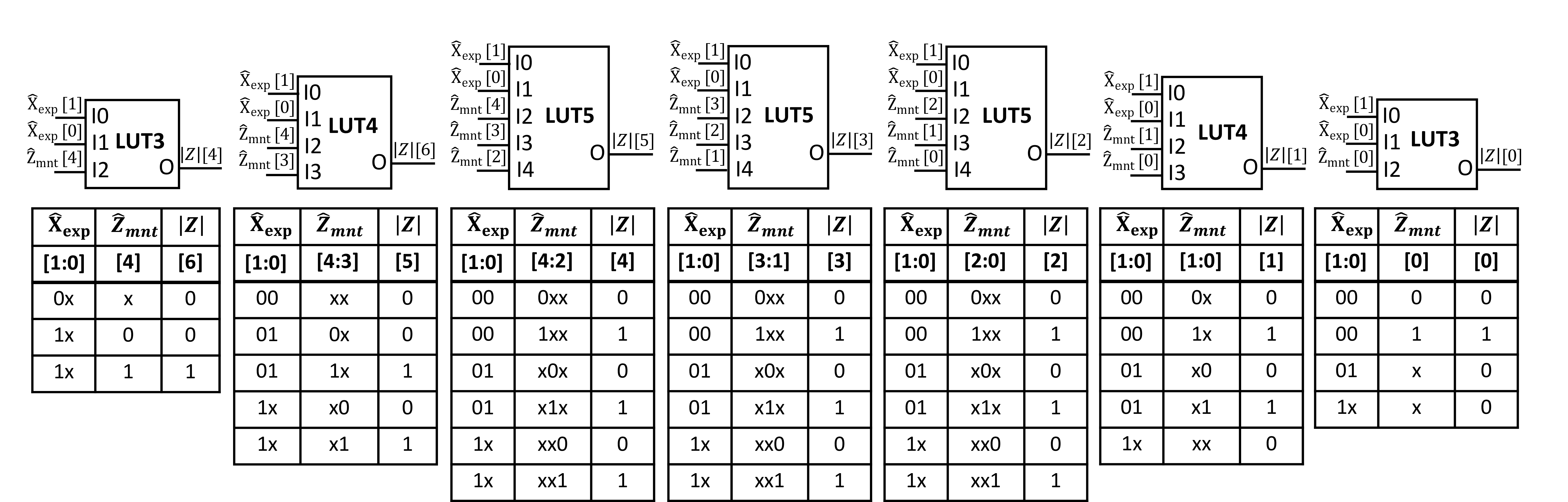}
    \caption{$float\left(1,2,5\right)$ to INT8 decoder, LUT mapping and corresponding truth tables.}
    \label{fig:4}
\end{figure*}

\textit{C) Floating-point to integer decoder—} The decoder converts the result of mantissa multiplication along with the exponent of the first operand into a $7$-bit unsigned format. The decoder needs seven LUT5 units, with each unit producing one output bit using two exponent bits and a maximum of three mantissa bits. The truth table of each output bit is shown in Fig. \ref{fig:4}. Since all LUT5 units function in parallel, the critical path is only one LUT5.

\textit{D) Two’s complement logic—} The output of the decoder is the absolute value of the result. As described in Section \ref{2.1}, the sign of the product can be obtained by exclusive OR operation between the sign bit of the two operands. If the product is negative, a two’s complement function must be applied. Two’s complement logic can be realized using eight LUT5 units. The carry signal is expected to be a major source of latency in this multiplier. To create an unsigned version of DyRecMul, we only need to remove the two’s complement stage, modify the floating-point format to float(0,2,5), and make slight adjustments to the encoder and decoder stages. The results in section \ref{3} indicate that the unsigned version is more efficient than the signed version. This is because the two's complement stage, which consumes a significant amount of hardware resources, is present only in the signed DyRecMul.

\section{Results}\label{3}
We evaluated INT8 DyRecMul by conducting error analysis, measuring hardware efficiency, and running accuracy tests on popular benchmark DL models. This section presents the results, discussions and comparisons.

\subsection{Error Analysis and Evaluation}\label{3.1}
One important method for assessing the accuracy of approximate calculations is error measurements and analysis. These assessments gauge the difference between the results of approximate calculations to those of exact calculations to determine any discrepancies \cite{vasicek2019formal}. This section showcases the accuracy of DyRecMul with five error metrics, Error Probability (EP), Mean Absolute Error (MAE), Mean Relative Error (MRE), Mean Squared Error (MSE), and Normalized Error Distance (NED):

\begin{equation}\label{eq:9}
    ED_{i} = \left |Exact_{i} - Approx_{i}\right|, i \in \mathbb{N}
\end{equation}

\begin{equation}\label{eq:10}
    EP=\frac{1}{2^{2N}}\sum_{i=0}^{2^{2N}-1}{ED}_i\neq0,
\end{equation}

\begin{equation}\label{eq:11}
    MAE=\frac{1}{2^{2N}}\sum_{i=0}^{2^{2N}-1}{ED}_i,
\end{equation}

\begin{equation}\label{eq:12}
    MRE=\frac{1}{2^{2N}}\sum_{i=0}^{2^{2N}-1}\frac{{ED}_i}{{Exact}_i},
\end{equation}

\begin{equation}\label{eq:13}
    MSE=\frac{1}{2^{2N}}\sum_{i=0}^{2^{2N}-1}{(ED_{i})^2},
\end{equation}

\begin{equation}\label{eq:14}
    NED=\frac{1}{2^{2N}}\sum_{i=0}^{2^{2N}-1}\frac{{ED}_i}{max(ED)},
\end{equation}
where $N$ denotes the multiplier bitwidth.

Table \ref{tab:table1} compares DyRecMul with existing approximated multipliers \cite{venkatachalam2017design,yang2017low,8017557,van2020fpga,liu2014low,danopoulos2022adapt,liu2017design,ansari2018low}. Our primary focus lies on the signed version of the DyRecMul, tailored specifically for machine learning (ML) applications. Nonetheless, we present the unsigned version of the multiplier, which is suitable for error-tolerant applications within the domain of image processing, such as image sharpening and smoothing \cite{jiang2018low}. Our method outperforms \cite{danopoulos2022adapt} by a significant margin in error analysis. The approximate multipliers proposed by \cite{van2020fpga} and \cite{liu2017design} offer reconfigurability based on an approximation factor $P$ representing the number of encoders employed in the partial product generation. DyRecMul yields comparable accuracy compared to this method in the low approximation $(P = 8)$ mode. In the mid-level approximation mode $(P = 10)$, our method outperforms the other designs and proves its robustness and effectiveness.

Although error analysis can be useful in evaluating the accuracy of approximated circuits, it may not give a complete picture. In section \ref{3.3}, we will examine how the approximation in DyRecMul affects the inference accuracy of some benchmark convolutional neural network (CNN) models in order to gain a more comprehensive understanding.

\begin{table}[t]
\renewcommand{\arraystretch}{1.5}
\large
\caption{ERROR ANALYSIS AND COMPARISONS}
\label{tab:table1}
\resizebox{\columnwidth}{!}{%
\begin{tabular}{|c|c|c|c|c|c|}
\hline
                           & \textbf{EP} & \textbf{MAE} & \textbf{MRE} & \textbf{MSE} & \textbf{NED} \\ \hline
DyRecMul (signed)          & 0.5157      & 397     & 0.0680       & 96336      & 0.00005   \\ \hline
Danopoulos {}\cite{danopoulos2022adapt}{}        & 0.7480      & 464     & 0.1259       & 5515991    & 0.01160       \\ \hline
Liu {}\cite{liu2017design}{} $(P = 8)$         & -           & -            & 0.0525       & -            & -            \\ \hline
Liu {}\cite{liu2017design}{} $(P = 10)$        & -           & -            & 0.1991       & -            & -            \\ \hline
Van Toan {}\cite{van2020fpga}{}   $(P = 8)$  & -           & 100          & 0.2299       & -            & -            \\ \hline
Van Toan {}\cite{van2020fpga}{}   $(P = 10)$ & -           & 512          & 0.9320       & -            & -            \\ \hline
DyRecMul (unsigned)        & 0.7380      & 336     & 0.0194       & 260528     & 0.1210       \\ \hline
Ha {}\cite{8017557}{}                & -           & 3490         & 0.3676       & -            & -            \\ \hline
Yang {}\cite{yang2017low}{}              & -           & 220          & 0.0196       & -            & -            \\ \hline
Venkatachalam {}\cite{venkatachalam2017design}{}     & -           & 101          & 0.0548       & -            & -            \\ \hline
Liu {}\cite{liu2014low}{}               & -           & 130          & 0.0062       & -            & -            \\ \hline
Ansari {}\cite{ansari2018low}{}            & -           & 1530         & 0.1336       & -            & -            \\ \hline
\end{tabular}%
}
\end{table}

\subsection{Hardware implementation results}\label{3.2}
DyRecMul was modelled in VHDL, then synthesized and implemented with AMD-Xilinx Vivado ML 2022.2 for a xcku5p Kintex Ultrascale+ FPGA with a speed grade of -$3$. For comparisons, we also evaluated the default AMD-Xilinx multiplier circuit. To ensure fairness, the synthesis tool was instructed to avoid using DSP resources.

Table \ref{tab:table2} presents the implementation costs and maximum supported clock frequency of DyRecMul and the Xilinx standard multiplier cores in three configurations: signed multiplier, unsigned multiplier and signed multiply and accumulation. The metrics include the utilization of LUTs and CARRY8 primitives, as well as the maximum clock frequency that is supported. Table II also presents the results of four unsigned approximate multipliers from existing works. These data have been extracted from the reported implementation results on Xilinx Spartan-6 in \cite{van2020fpga}, and normalized based on exact multiplier results.  In all the examined designs, the utilization of other resources such as BRAMs and DSPs is zero. The results indicate significant reductions in LUT usage with 64\%, 80\%, and 67\% savings for the signed, unsigned, and MAC cases, respectively, compared to the exact $8\times8$ multiplier. DyRecMul also archived a higher frequency than all existing exact and approximate multipliers.  

Despite considerable utilization of CARRY8 fast carry primitives in standard multipliers, DyRecMul achieves a significantly higher maximum clock frequency thanks to its optimized datapath. DyRecMul yields similar results in both signed multiplier and MAC setups. This is because the two's complement logic at the end of the architecture needs the same amount of LUTs as an adder/subtractor in MAC. In other words, an adder/subtractor can be incorporated into the two’s complement LUT-based circuit with almost no cost overhead. Among the three tested setups, DyRecMul achieves the highest performance in the unsigned multiplication setup. This is because the two’s complement logic in signed multiplication and the add/sub logic in MAC operations constitute a significant portion of both hardware and latency.

\begin{table}[h]
\renewcommand{\arraystretch}{1.5}
\large
\caption{IMPLEMENTATION COSTS AND PERFORMANCE ON A KINTEX ULTRASCALE+ }
\label{tab:table2}
\resizebox{\columnwidth}{!}{%
\begin{tabular}{|c|c|c|c|c|c|}
\hline
\textbf{Function}                    &                             & \textbf{Size} & \textbf{\#LUT} & \textbf{\#CARRY8} & \textbf{Frequency (MHz)} \\ \hline
\multirow{3}{*}{Signed Multiplier}   & DyRecMul                    & 8×8           & 25             & 0                 & 770                      \\ \cline{2-6} 
                                     & \multirow{2}{*}{\begin{tabular}[c]{@{}c@{}}AMD-Xilinx \\ (Exact)\end{tabular}} & 8×8           & 69             & 8                 & 730                      \\ \cline{3-6} 
                                     &                             & 7×7           & 61             & 6                 & 660                      \\ \hline
\multirow{3}{*}{Signed MAC}          & DyRecMul                    & 8×8           & 25             & 1                 & 769                      \\ \cline{2-6} 
                                     & \multirow{2}{*}{\begin{tabular}[c]{@{}c@{}}AMD-Xilinx \\ (Exact)\end{tabular}} & 8×8           & 76             & 10                & 571                      \\ \cline{3-6} 
                                     &                             & 7×7           & 69             & 8                 & 585                      \\ \hline
\multirow{7}{*}{Unsigned Multiplier} & DyRecMul                    & 8×8           & 16             & 0                 & 950                      \\ \cline{2-6} 
                                     & \multirow{2}{*}{\begin{tabular}[c]{@{}c@{}}AMD-Xilinx \\ (Exact)\end{tabular}} & 8×8           & 82             & 6                 & 684                      \\ \cline{3-6} 
                                     &                             & 7×7           & 55             & 6                 & 725                      
                                     \\ \cline{2-6}
                                     & \multirow{1}{*}{Venkatachalam {}\cite{venkatachalam2017design}{}} & 8×8           & 108             & 2                 & 781                   
                                     \\ \cline{2-6}
                                     & \multirow{1}{*}{Ha {}\cite{8017557}{}} & 8×8           & 73             & 2                 & 751                              
                                     \\ \cline{2-6}
                                     & \multirow{1}{*}{Ansari {}\cite{ansari2018low}{}} & 8×8       & 62             & 2                 & 856                          
                                     \\ \cline{2-6}
                                     & \multirow{1}{*}{Yang {}\cite{yang2018low}{}} & 8×8         & 76             & 2                 & 820                           \\ \hline

\end{tabular}%
}
\end{table}

\begin{figure*}[h]
    \centering
    \includegraphics[width = 460pt]{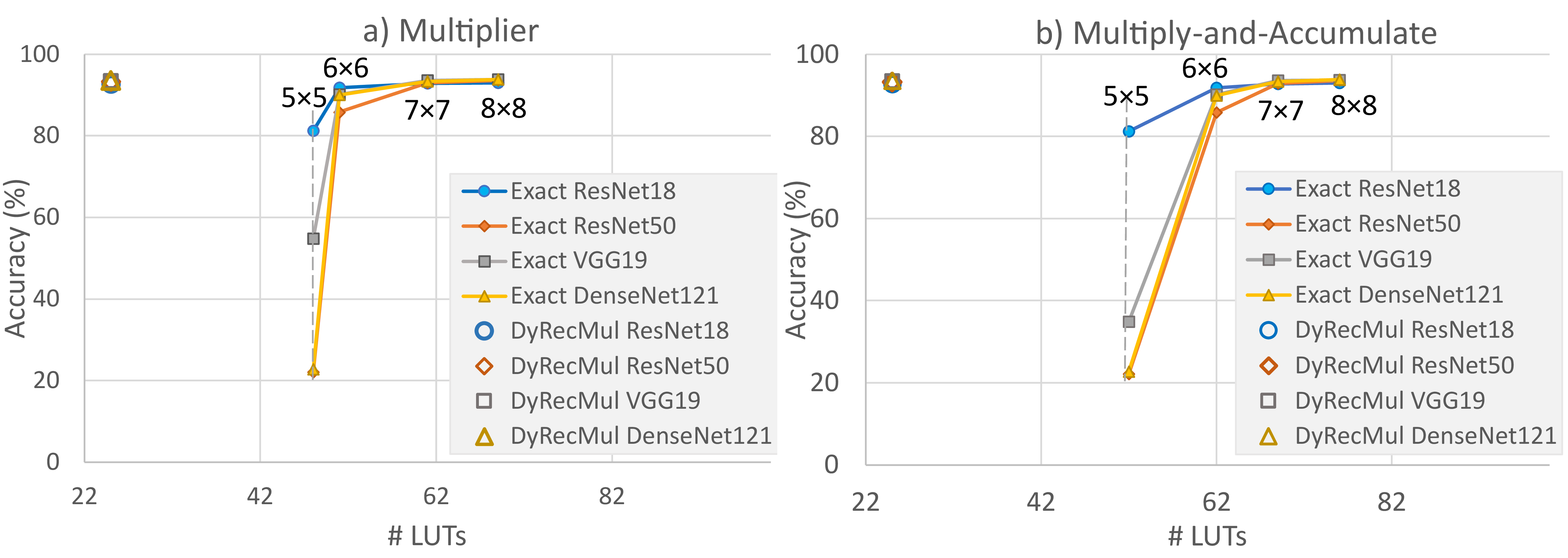}
    \caption{Trade-offs between hardware utilization and inference accuracy when running four benchmark CNNs using DyRecMul and standard exact multipliers of different sizes. (a) signed multiplier. (b) MAC.}
    \label{fig:5}
\end{figure*}

\begin{table*}[h]
\renewcommand{\arraystretch}{1.1}
\caption{INFERENCE ACCURACY AND HARDWARE METRICS COMPARISON WITH PREVIOUS STUDIES}
\label{tab:table3}
\resizebox{\textwidth}{!}{%
\begin{tabular}{|c|c|cccc|cc|cc|}
\hline
\textbf{Multiplier}   & \textbf{Dataset} & \multicolumn{4}{c|}{\textbf{Accuracy (\%)}}                                                                                                          & \multicolumn{2}{c|}{\textbf{Area (LUTs)}}        & \multicolumn{2}{c|}{\textbf{Max Freq. (MHz)}}    \\ \hline
                      &                  & \multicolumn{1}{c|}{\textbf{ResNet18}} & \multicolumn{1}{c|}{\textbf{ResNet50}} & \multicolumn{1}{c|}{\textbf{VGG19}} & \textbf{DenseNet121} & \multicolumn{1}{c|}{\textbf{MUL}} & \textbf{MAC} & \multicolumn{1}{c|}{\textbf{MUL}} & \textbf{MAC} \\ \hline
Exact    $(5\times5)$        & CIFAR-10         & \multicolumn{1}{c|}{81.20}            & \multicolumn{1}{c|}{22.08}              & \multicolumn{1}{c|}{34.83}         & 22.72              & \multicolumn{1}{c|}{48}           & 52           & \multicolumn{1}{c|}{740}          & 724          \\ \hline
Exact    $(6\times6)$        & CIFAR-10         & \multicolumn{1}{c|}{91.83}            & \multicolumn{1}{c|}{85.81}            & \multicolumn{1}{c|}{90.17}         & 89.98               & \multicolumn{1}{c|}{51}           & 62           & \multicolumn{1}{c|}{770}          & 700          \\ \hline
Exact    $(7\times7)$        & CIFAR-10         & \multicolumn{1}{c|}{92.83}            & \multicolumn{1}{c|}{92.95}            & \multicolumn{1}{c|}{93.60}         & 93.37              & \multicolumn{1}{c|}{61}           & 69           & \multicolumn{1}{c|}{660}          & 585          \\ \hline
Exact $(8\times8)$           & CIFAR-10         & \multicolumn{1}{c|}{92.98}            & \multicolumn{1}{c|}{93.57}            & \multicolumn{1}{c|}{93.78}         & 93.85              & \multicolumn{1}{c|}{69}           & 76           & \multicolumn{1}{c|}{73}          & 571          \\ \hline
{}\cite{danopoulos2022adapt}{} $(8\times8)$   & CIFAR-10         & \multicolumn{1}{c|}{-}                  & \multicolumn{1}{c|}{82.70}              & \multicolumn{1}{c|}{90.70}            & -                    & \multicolumn{1}{c|}{-}            & -            & \multicolumn{1}{c|}{-}            & -            \\ \hline
{}\cite{hammad2021cnn}{}   SSM $(8\times8)$ & ImageNetV2       & \multicolumn{1}{c|}{-}                  & \multicolumn{1}{c|}{-}                  & \multicolumn{1}{c|}{92.28}           & -                    & \multicolumn{1}{c|}{-}            & -            & \multicolumn{1}{c|}{-}            & -            \\ \hline
{}\cite{hammad2021cnn}{}   DSM $(8\times8)$ & ImageNetV2       & \multicolumn{1}{c|}{-}                  & \multicolumn{1}{c|}{-}                  & \multicolumn{1}{c|}{92.15}           & -                    & \multicolumn{1}{c|}{-}            & -            & \multicolumn{1}{c|}{-}            & -            \\ \hline
DyRecMul              & CIFAR-10         & \multicolumn{1}{c|}{92.72}            & \multicolumn{1}{c|}{93.32}            & \multicolumn{1}{c|}{93.45}         & 93.54              & \multicolumn{1}{c|}{25}           & 25           & \multicolumn{1}{c|}{770}          & 769          \\ \hline
\end{tabular}%
}
\end{table*}

\subsection{Accuracy in deep learning computation}\label{3.3}
To ensure the usability of DyRecMul in DNN accelerators, we evaluated its accuracy for the benchmark models ResNet18, ResNet50, VGG19, and DenseNet121 using the CIFAR-10 dataset. For these experiments, we used the AdaPT framework which provides a rapid emulation environment to measure the accuracy of new approximate multipliers in the CNNs, LSTMs, and GANs inferences \cite{danopoulos2022adapt}. Fig. \ref{fig:5} compares the inference accuracy and the hardware utilization costs offered by an INT8 DyRecMul with those of standard AMD-Xilinx multipliers with different bitwidths. The accuracy is measured for post-training deployment with no re-training applied. Fig. \ref{fig:5} clearly indicates that DyRecMul offers a significantly superior accuracy-hardware cost trade-off compared to the standard AMD-Xilinx multiplier cores. Table \ref{tab:table3} provides detailed results of DyRecMul and selected previous works \cite{hammad2021cnn}, including inference accuracy, maximum supported clock frequency and hardware utilization. Based on the findings, DyRecMul provides an average accuracy that is only 0.29\% lower than that of $8\times8$ exact multipliers and 0.07\% higher than $7\times7$. The worst-case accuracy distance from the exact $8\times8$ multiplier is only 0.33\%. Table \ref{tab:table3} also shows that DyRecMul offers slightly higher accuracy compared to Danopoulos \cite{danopoulos2022adapt}, SSM, and DSM \cite{hammad2021cnn} INT8 approximate multipliers in the VGG19 test and significantly higher accuracy compared to Danopoulos in ResNet50. Also, as reported in Section III-A, DyRecMul uses fewer LUTs, with a reduction of 64\% and 67\% in signed and MAC setups, respectively. Furthermore, it can support clock frequencies that are up to 5.5\% and 34.6\% higher than an exact $8\times8$ multiplier in signed and MAC setups, respectively.

\section{Conclusion}\label{4}
This paper introduces DyRecMul, an approximate multiplier meticulously optimized for machine learning computations on FPGAs. Leveraging dynamically reconfigurable logic, this multiplier achieves remarkable hardware efficiency. Additionally, it employs a cost-effective internal floating-point conversion technique to preserve a wide dynamic range, thereby enhancing the precision of machine learning calculations.
The results demonstrate that for INT8 precision, DyRecMul requires 64\%, 80\%, and 67\% fewer LUTs compared to the Xilinx standard multiplier core in signed, unsigned, and MAC setups, respectively. Moreover, the maximum supported clock frequency remains notably higher than that of Xilinx multipliers.
DyRecMul also provides a substantial advantage over four existing approximate multipliers in terms of both hardware utilization and frequency. Additionally, the results indicate that employing this multiplier for post-training deep learning inference leads to a minimal average accuracy degradation of less than 0.29\% compared to exact INT8 multiplication.

\bibliographystyle{IEEEtran}
\bibliography{References}

\end{document}